\documentclass{jetpl}

\usepackage[cp1251]{inputenc}
\usepackage[english,russian]{babel}
\twocolumn

\title{Quark cluster contribution to cumulative proton emission
in fragmentation of carbon ions.}

\rtitle{Quark cluster effect\ldots}

\author{B.\,M.\,Abramov, P.\,N.\,Alekseev, Yu.\,A.\,Borodin,
S.\,A.\,Bulychjov, I.\,A.\,Dukhovskoy, A.\,I.\,Khanov,
A.\,P.\,Krutenkova\/\thanks{e-mail: anna.krutenkova@itep.ru},
V.\,V.\,Kulikov, M.\,A.\,Martemyanov, M.\,A.\,Matsyuk,
E.\,N.\,Turdakina}

\address{Institute for Theoretical and Experimental Physics, 117218 Moscow, Russia}

\abstract{In the FRAGM experiment at heavy ion accelerator complex
TWAC-ITEP, the proton yields at an angle 3.5$^\circ$ have been measured
at fragmentation of carbon ions at  $T_0 = $ 0.6, 0.95 and 2.0
GeV/nucleon on beryllium target. The data are presented as
invariant proton yields on cumulative variable $x$ in the range
0.9 < $x$ < 2.4. Proton spectra cover six orders of invariant
cross section magnitude. They have been analyzed in the framework
of quark cluster fragmentation model. Fragmentation functions of
quark-gluon string model are used. The probabilities of the
existence of multi-quark clusters in carbon nuclei are estimated
to be 8--12\% for six-quark clusters and 0.2--0.6\% for nine-quark
clusters.  }

\addto\captionsrussian{\def\refname{}}
\begin{document}

\maketitle

{\bf Introduction}. Since the discovery of cumulative effect  in
seventies \cite{baldin,leksin}, the question of the nature of
cumulative particles  is still under discussion.
Cumulative particles are produced in interactions with nuclei in kinematic
region forbidden for interaction with free nucleon. Few hypothesis
of their origin have been considered. Among them there are fluctuations
of nuclear matter \cite{flucton}, clusters \cite{efremov}, few-nucleon correlations
\cite{strikman}, excited dibaryons \cite{kukulin}, etc.  It was pointed out
\cite{kopel, braun} that multiple scattering in nuclear matter
can also contribute. Modern theoretical approaches (see, e.g.,
\cite{krivoruchenko}) connect the cumulative effect with
contribution from multi-quark states. These states (first of all,
six-quark states) are considered for nuclear matter phase
transitions at high densities. At intermediate energies
\cite{pirner}, the existence of multi-quark clusters in cold and
hot baryonic matter blurs the boundary between hadronic and
quark-gluon phases. Probability of the two-nucleon fluctuation
in $^{12}$C nucleus was for the first time estimated in
\cite{burov} based on the experimental data on cumulative pion
production. It is in qualitative
agreement with theoretical prediction, which was obtained later within
quark cluster model \cite{sato}. Similar approach, in which
fragmentation functions of quark clusters were calculated within
the quark-gluon string model, was successfully used in
\cite{efremov} for the description of $K^{-}$, $\pi^{-}$ and
antiproton inclusive spectra in hadron-nucleus interactions.

Ion beams open new ways for the cumulative effect study.
One of the authors of \cite{efremov}, A.B. Kaidalov,
suggested to use the data from our experiment FRAGM  \cite{fragm,
abramov} for the experimental estimation of the admixture of the
multi-quark cluster state in nuclear matter.  In this experiment,
cumulative protons are measured in inverse kinematics, i.e. in the
fragmentation region of projectile nucleus. Such measurement has
definite advantages over measurements in the target fragmentation
region. First, relativistic boost of forward going protons
increases considerably proton detection solid angle in the
rest-frame of fragmented nucleus at fixed acceptance of a detection system in
laboratory frame. Second, in the inverse kinematics there are no
problems with a detection of protons which are at rest in
projectile rest frame because they have lab momentum close to
momentum per nucleon of the projectile. This allows to clearly observe
nucleon-nucleon component of the projectile and use it for normalization
that is impossible for
target fragmentation. However inverse kinematic creates additional
problems, connected with measurements of the protons with momentum
few times higher than momentum per nucleon of the projectile.

{\bf Experiment}. In the FRAGM experiment at the accelerator
complex TWAC (Tera-Watt Accumulator) ITEP, the yields of nuclear
fragments are studied in carbon fragmentation on beryllium
target:
\begin{equation}
^{12}\mathrm{C} + \mathrm{Be}  \rightarrow \mathrm{f + X }.
\end{equation}
The main goal of the experiment is to collect data at high momenta
of nuclear fragments f. In this paper the proton spectra from the
reaction (1), obtained at carbon kinetic energies $T_0$ = 0.6,
0.95 and 2.0 GeV/nucleon are analyzed. A measurement of momentum
spectra at different energies gives a possibility to study energy
dependence of their parameters.

\begin{figure}[h!]
\vspace{0.2cm}
\includegraphics[width=0.45\textwidth]{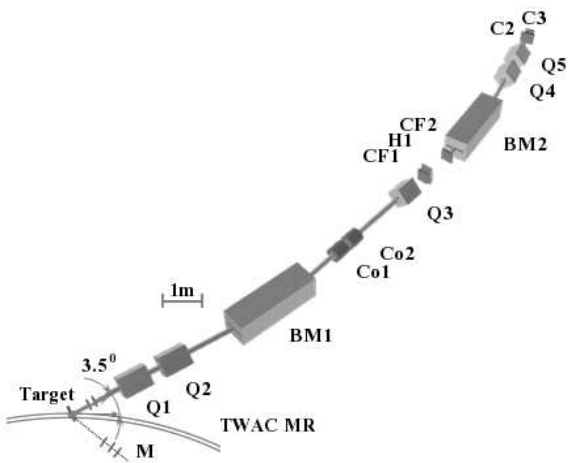}
\caption {Fig.1. Set-up of the FRAGM experiment.
TWAC MR is the TWAC main ring, М is a monitor, Q1-Q5 are
quadrupoles, BM1 and BM2 are bending magnets, Co1 and Co2 are
collimators, CF1, CF2, H1, C2 and C3 are scintillator counters.}
\end{figure}
Experimental set-up (Fig. 1) is comprised of two-step magneto-optical
channel, placed at an angle   $\theta$ = 3.5$^\circ$ with respect
to internal ion beam of the ITEP accelerator. Narrow vertical
strip of 200 $\mu m$ Be-foil was used as a target which allows to
have simultaneously both high luminosity due to multiple passage
of ions through the target and small sizes of the source for full
usage of high momentum resolution of the channel. The first step
of the channel consists of quadrupole doublet  Q1, Q2, bending
magnet BM1 and achromatic correction quadrupole Q3, placed in
the first focus of the channel. The second step contains bending
magnet BM2 and quadrupole doublet Q4, Q5. This step refocuses the
target image from the first focus to the second one, sixteen
meters downstream. Scintillator hodoscope of twenty
vertical and eight horizontal elements with sizes 20х1х1 cm$^{3}$
for beam profile measurements is placed in the first focus. This
hodoscope allows to improve fragment momentum resolution up to
0.2\%, using focusing properties of magneto-optical channel with
full momentum acceptance of $\pm$3\%. Scintillator counters CF1,
CF2 and С2, С3 for amplitude and time-of-flight
(TOF) measurements are placed in the first and in the second focuses.
Each counter is seen by two PMT's from the opposite sides to
use their signals in mean-time determination.
Coincidence of signals from the counters of the first and
the second focuses gives a trigger, which starts  the data
transfer from CAMAC system to a computer under
LINUX. Readout software is based on ROOT \cite{root} package. As a
monitor М, a telescope of three scintillator counters, which view
the target at an angle of about 2$^\circ$ is used.

Protons are selected on correlation plot of signal charge from a
counter (a function of fragment charge) $vs$ TOF (a function
of fragment mass). As an example, correlation plot for
projectile energy of 0.6 GeV and momentum of magneto-optical
channel of 2.15 GeV/c is shown in Fig.2. Groups of events
corresponding to outgoing fragments: hydrogen, helium, lithium,
beryllium, boron and carbon isotopes are clearly seen.
\begin{figure}[h!]
\vspace{0.2cm}
\includegraphics[width=0.45\textwidth]{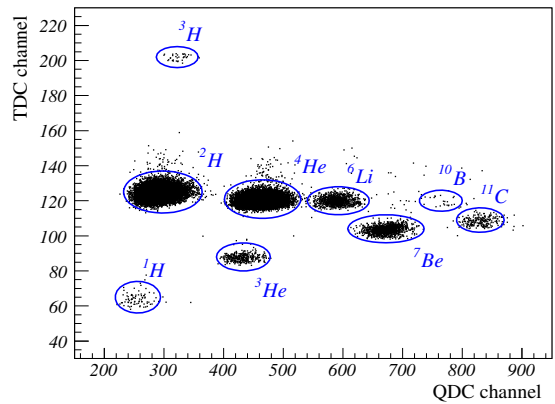}
\caption {Fig.2 Correlation plot of signal charge from one of
scintillation counters (QDC) $vs$ time-of-flight (TDC) at a base
of 16 m. Projectile energy is 600 MeV/nucleon, fragment rigidity is
2.15 GeV/c.}
\end{figure}
At each projectile energy the proton yields are measured by momentum scan
 of the channel with a step 25-200 MeV
c$^{-1}$. Set-up efficiency is calculated with  GEANT4-based
\cite{geant1,geant2} simulation program. The program traces
particles in the magneto-optical channel taking into account
multiple scattering effects, ionization losses and absorption in
the detector materials.

{\bf Data analysis}. The invariant cross sections of proton yield
$\sigma_{inv} = (E/p_{0}) d^2\sigma/dx d(p_t^2)$ as a function of
cumulative variable $x = p/p_0$ for projectile energies $T_0$
=0.6, 0.95 and 2.0 GeV/nucleon are shown in Fig.3 - 5. Here $p_0$
is projectile momentum per nucleon, $p$ is proton momentum in the
laboratory frame, $p_t$ is its transverse component with respect
to projectile. The data cover six orders of invariant cross section
magnitude. It
is three orders of magnitude more than in the most sensitive
previous experiment \cite{anderson} at 1.05 GeV/nucleon.
\begin{figure}[h!]
\vspace{0.2cm}
\includegraphics[width=0.45\textwidth]{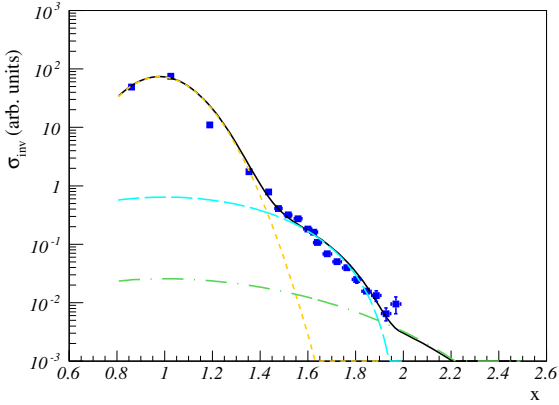}
\caption {Fig.3. Points are invariant cross section $\sigma_{inv}$
of proton yield at projectile energy of 0.6 GeV/nucleons in
arbitrary units as function of $x$ variable. Solid curve is a fit
to the data in quark-cluster model. Contributions from one-, two-
and three-nucleon clusters are shown with dashed, dotted and
dash-dotted lines, respectively.}
\end{figure}
\begin{figure}[h!]
\vspace{0.2cm}
\includegraphics[width=0.45\textwidth]{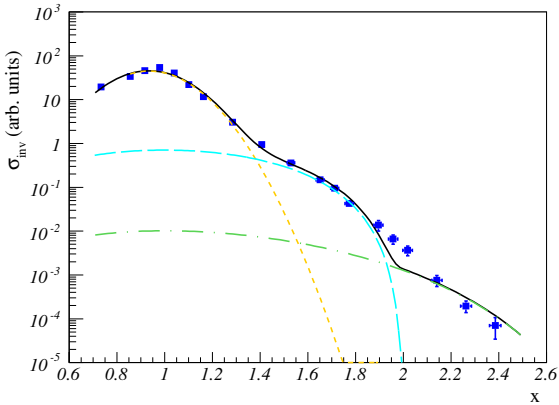}
\caption {Fig.4. The same as Fig.3, but at 0.95 GeV/nucleon.}
\end{figure}
\begin{figure}[h!]
\vspace{0.2cm}
\includegraphics[width=0.45\textwidth]{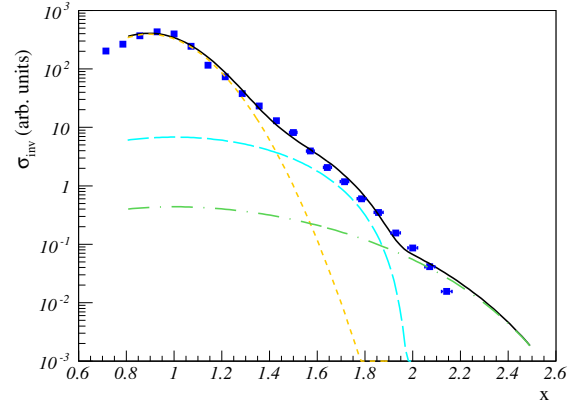}
\caption {Fig.5. The same as Fig.3, but at 2.0 GeV/nucleon.}
\end{figure}
In the region of the maximum ($x\approx$ 1.0) the shapes of the
spectra are close to Gaussians as predicted by statistical models.
However, already at $x\geq$ 1.3 the spectra become exponentials,
which is typical for cumulative processes.

As was already mentioned, the most successful approach to the
problem of cumulative particle production is the quark cluster
model \cite{efremov}. This model was used to describe yields of
cumulative pions, kaons and antiprotons. However, for protons such
analysis was not performed. Reliable identification of one-nucleon
component achievable in inverse kinematics allows to apply this analysis also
for protons.  In the framework of this model clusters existing in
a nucleus, consist of 3$k$ ($k$ = 1, 2, 3,...) valent quarks.
Conventional nucleon component of the nucleus corresponds to $k$ =
1. The probabilities $w_k$ of such clusters
in a nucleus with A nucleons are normalized by
$\sum\limits_{k=1}^{A-1} w_k$ = 1. As cluster contribution into
observed processes falls with increasing  $k$, we limit ourselves
with $k$ = 1, 2, 3 only and represent the invariant cross section
as a sum of three components:
\begin{equation}
\sigma_{inv} \propto Gw_1g(x,p_t^2) + w_2 b_2(x,p_t^2) + w_3
b_3(x,p_t^2),
\end{equation}
where functions $g$, $b_2$ and $b_3$ are fragmentation functions of
quark clusters into protons. $g$ is a gaussian
\begin{equation}
g(x,p_t^2)\hspace{-0.08cm}=\hspace{-0.08cm}\exp(\hspace{-0.05cm}- 0.5 ((1\hspace{-0.02cm}-\hspace{-0.02cm}\Delta)\hspace{-0.02cm}-\hspace{-0.02cm}x)^2\hspace{-0.05cm}/\hspace{-0.05cm}\sigma_x^2 )\exp(\hspace{-0.05cm}-0.5 p_t^2\hspace{-0.05cm}/\hspace{-0.05cm} \sigma_p^2),
\end{equation}
$b_2$ and $b_3$ are calculated in the framework of the quark-gluon
string model:
\begin{equation}
b_2(x,p_t^2) \hspace{-0.05cm}=\hspace{-0.05cm} \left\{\begin{array}{l}
\hspace{-0.1cm}B_2 (x/2)^3(1\hspace{-0.05cm}-\hspace{-0.05cm}x/2)^3\exp(-\alpha_1p_t^2),\\
\hspace{-0.1cm}0,  ~ x \notin [0,2],
\end{array}\right.\hspace{-0.15cm}
\end{equation}
\begin{equation}
b_3(x,p_t^2)\hspace{-0.05cm}=\hspace{-0.05cm} \left\{\begin{array}{l}
\hspace{-0.1cm}B_3 (x/3)^3(1\hspace{-0.05cm}-\hspace{-0.05cm}x/3)^6\exp(-\alpha_2p_t^2),\\
\hspace{-0.1cm}0,  ~ x \notin [0,3].
\end{array}\right.\hspace{-0.15cm}
\end{equation}

The values $G, B_2$ and $B_3$ are known normalization constants;
the first one is defined by Gaussian normalization
\begin{equation}
G = (4 \sqrt{2\pi} \sigma_x \sigma_p^2)^{-1},~~\sigma_p = \sigma_x m_pp_0/(T_0+m_p),
\end{equation}
and two others, $B_2$ and $B_3$, can be found from:
\begin{equation}
\int\limits_0^\infty \int\limits_0^\infty b_i(x,p_t^2)dxdp_t^2=i/2, ~ i=2,3.
\end{equation}
Transverse momentum dependence is not predicted within quark gluon string model.
We have got the invariant cross section slopes on $p_t^2$ from the data
\cite{anderson}  and extrapolated them to the region of our
experiment, the values  $\alpha_1$ = 5
GeV$^{-2} c^2$ and $\alpha_2$ = 3 GeV$^{-2} c^2$ were used.

The results of the data fit with the formula
(2) at projectile energies 0.6, 0.95 and 2.0 GeV are shown in Fig.
3 - 5. The fitted curve is shown by solid line. The
contributions of one-nucleon (3q) component, two-nucleon (6q) and
three-nucleon (9q) clusters are given by dashed, dotted and
dashed-dotted lines, respectively. The fit parameters are mean
value (1-$\Delta$) and r.m.s. of Gaussian ($\sigma_{x}$), and
also probabilities $w_2$ and $w_3$, connected to $w_1$ with the
relation $w_1 + w_2 +w_3$ =1. The obtained probabilities  $w_2$
and $w_3$ for carbon nucleus are
given in the Table.
\begin{table}[bhpt]
\caption{{\normalsize Table}. Results from the fits of Fig.3 - 5.
$T_0$ and $p_0$ are kinetic energy and momentum of the projectile
per nucleon, $x_{max}$ is maximal  value of $x$, reached in this
experiment,
probabilities $w_2$ and $w_3$ for carbon nucleus are defined in
the text. Statistical errors of the fit are
given in parentheses.}
\begin{tabular}{|c|c|c|c|c|}
\hline
$T_0$,& $p_0$,& $x_{max}$&  $w_2$ & $w_3$  \\
GeV/n & GeV/c/n & & &    \\
\hline
0.6 &1.22 & 1.95 &  .077(10) & .004(2) \\
\hline
.95 &1.6 & 2.4  & .119(17) & .002(1)  \\
\hline
2.0 & 2.72 & 2.15 & .098(18) & .006(1)  \\
\hline
\end{tabular}

\end{table}
The two-nucleon cluster probability, estimated at
different projectile energies varies within 7.7 - 11.9\%, while
the three-nucleon one is within 0.2 - 0.6\%. They are
compatible both with given statistical errors of the fit and with
expected independence of these probabilities on projectile energy.
It should be emphasized that obtained probabilities for carbon
nucleus could be considered only as
estimates because of difficulties to take into account
systematic uncertainties of the theoretical approach. First of all
the fragmentation functions (4) and (5) were justified in the
quark-gluon string model only in the
boundary regions near $x$=0, $x$=2 and $x$=3. The calculation of
fragmentation functions in the whole $x$ range is still unsolved
problem of the model. The fragmentation functions used here are the
simplest (and widely used) ones, satisfying above
mentioned boundary conditions. Moreover, in Fig.3 - 5 at $x
> $ 1.3 one can see wave-like behavior of the resulting fit curve
while the data demonstrate purely exponential fall off with $x$.

This can be considered as indication that the model does not take
into account some small effects which smear fragmentation
functions behavior. One of them is evident, it is
internal motion of two- and three-nucleon clusters in a nucleus
which has not been taken into account in this approach. It can be expected that
improvement of multi-quark cluster approach will allow to overcome
these difficulties. The important result of the presented analysis
is the demonstration of possibility to use data on cumulative
proton yield in the inverse kinematics for estimates of these probabilities.
Obtained value of $w_2$ is close to the value of 6\%, obtained
in \cite{burov} from cumulative pion production and to
theoretical predictions of 12.5\% given in \cite{sato}.
However, the value of $w_3$ is much smaller than  2.6\% predicted
in \cite{sato}. The values $w_2$ ($w_3$) are not far from
probabilities of  two-nucleon (three-nucleon)
correlations in nuclei, obtained in the TJNAF experiment on
A(e,e') inclusive  electron scattering on nuclei \cite{egiyan},
(19.3 $\pm$ 4.1)\% and (0.55 $\pm$ 0.17)\% for carbon nucleus.
Reasonable agreement of the results of these experiments
can be considered as evidence for unique nature of
quark clusters and short-range nucleon correlations in nuclei.

Authors would like to thank K.G. Boreskov, O.V. Kancheli, Yu.T.
Kiselev, V.K. Lukyanov and I.I. Tsukerman for fruitful
discussions. We are also indebted to the personnel of TWAC-ITEP and
technical staff of the FRAGM experiment. The work has been
performed with financial support of the RFBR (grant
№12-02-01111а).

\end{document}